\documentclass[aps,twocolumn,showpacs, showkeys,nofootinbib,floatfix]{revtex4-1}

\usepackage{amssymb}
\usepackage{amsmath}
\usepackage{graphicx}
\usepackage{hyperref}
\usepackage{longtable}

\begin{document}

\title{Unified Dark Fluid with Constant Adiabatic Sound Speed and Cosmic Constraints}

\author{Lixin Xu$^{1,2,3}$}
\email{lxxu@dlut.edu.cn}
\author{Yuting Wang$^{2,4}$}
\email{yuting.wang@port.ac.uk}
\author{Hyerim Noh$^1$}
\email{hr@kasi.re.kr}

\affiliation{$^1$Korea Astronomy and Space Science Institute,
Yuseong Daedeokdaero 776,
Daejeon 305-348,
R. Korea}
\affiliation{$^2$Institute of Theoretical Physics, School of Physics \&
Optoelectronic Technology, Dalian University of Technology, Dalian,
116024, P. R. China}

\affiliation{$^3$College of Advanced Science \& Technology, 
Dalian University of Technology, Dalian, 116024, P. R. China}

\affiliation{$^4$Institute of Cosmology \& Gravitation, 
University of Portsmouth, Portsmouth, PO1 3FX, United Kingdom}

\begin{abstract}

As is known above $90\%$ of the energy content in Universe is made of unknown dark component. Usually this dark fluid is separated into two parts: dark matter and dark energy. However, it may be a mixture of these two  energy components, or just one exotic unknown fluid. This property is dubbed as dark degeneracy. With this motivation, in this paper, a unified dark fluid having constant adiabatic sound speed $c_s^2=\alpha$, which is in the range $[0,1]$, is studied.  At first, via the energy conservation equation, its energy density, $\rho_d/\rho_{d0}=(1-B_s)+B_s a^{-3(1+\alpha)}$ where $B_s$ is related to integration constant from energy conservation equation as another model parameter, is presented. Then by using Markov Chain Monte Carlo method with currently available cosmic observational data sets which include type Ia supernova Union 2, baryon acoustic oscillation and WMAP $7$-year data of cosmic background radiation, we show that small values of $\alpha$ are favored in this unified dark fluid model. Furthermore, we show that smaller values of $\alpha<10^{-5}$ are required to match matter (baryon) power spectrum from SDSS DR7.
\end{abstract}



\maketitle

\section{Introduction}

Since the discovery of an accelerated expansion of our universe \cite{ref:Riess98,ref:Perlmuter99}, a flood of dark energy models have been presented to describe an exotic energy component which has negative pressure and pushes the universe to an accelerated expansion. For recent reviews on dark energy, please see \cite{ref:DEReview1,ref:DEReview2,ref:DEReview3,ref:DEReview4,ref:DEReview5,ref:DEReview6,ref:DEReview7}. As is known above $90\%$ of the energy content in Universe is made of unknown dark component. However, the nature of this dark fluid is still unknown. Usually, this dark fluid is divided into two parts: dark matter and dark energy. But as pointed out by the authors \cite{ref:darkdeneracy,ref:DG}, there exists another possibility that it is a mixture of dark matter and dark energy, or just one exotic unknown fluid. This property is dubbed as dark degeneracy. A unified dark fluid is defined from the Einstein field equation
\begin{equation}
T^{dark}_{\mu\nu}=\frac{1}{8\pi G}G_{\mu\nu}-T^{obs}_{\mu\nu},
\end{equation} 
where $T^{obs}_{\mu\nu}$ is the observed energy components and $G_{\mu\nu}$ is the observed geometric structure of Universe. In the framework of linear perturbation theory, the micro scale properties of a fluid are characterized by its equation of state (EoS) $w_d$ and sound speed $c_s^2$. As a contrast to the reports where a parametrized form of EoS of dark energy is assumed in the so-called model-independent formalism, we study a unified dark fluid with constant adiabatic sound speed $c_s^2=\alpha$. Actually, the case of zero adiabatic sound speed has been studied in Ref. \cite{ref:DG,ref:cs0}. And time variable sound speed cases were also discussed in \cite{ref:csvar}. In this paper, we consider a generalized case, i.e $c_s^2=\alpha$. The similar situation was also discussed in \cite{ref:lalphacdm} where an effective cosmological constant $\rho_\Lambda$ and dark matter were defined. And it was dubbed as $\Lambda\alpha$CDM model. In fact, for a unified dark fluid, one can do different decomposition, for example defining dark matter $\rho_{dm}\equiv\rho_{dm0}a^{-3}$ and the remaining part as dark energy $\rho_{de}\equiv\rho_d-\rho_{dm}$. However, without the guide of physics principle, any decomposition would be non-proper. Moreover, it may be just an entirely whole energy component. And there does not exist any decomposition at all. So as a contrast to Ref. \cite{ref:lalphacdm} , in this paper, we prefer taking it as an entirely whole dark fluid and show the properties of dark degeneracy. Furthermore instead of using the location of CMB acoustic peaks as done in \cite{ref:lalphacdm}, the full information from WMAP $7$-year data sets will be used in this paper. As a result, a tighter constraint will be obtained.

At first, in section \ref{sec:DG}, the energy density and background evolution equation for a unified dark fluid of constant adiabatic sound speed are shown. The corresponding scalar fields action is also presented. In section \ref{sec:method}, by using Markov Chain Monte Carlo (MCMC) method with currently available cosmic observational data sets which include type Ia supernova Union 2, baryon acoustic oscillation and WMAP $7$-year CMB, we show the parameter space. A summary is presented in section \ref{ref:conclusion}.

\section{Generalized Dark Degeneracy Fluid with a Constant Adiabatic Sound Speed}   \label{sec:DG}  

\subsection{Background Equations, Perturbations and Instabilities}

Consider a unified dark fluid as a barotropic fluid with constant adiabatic sound speed 
\begin{equation}
c_s^2=\alpha.
\end{equation}
For barotropic fluid the adiabatic sound speed equals the speed of which perturbations propagate in the fluid. Generally its EoS is given as
\begin{equation}
P(\rho)=w(\rho)\rho,
\end{equation}
where $w(\rho)$ is the EoS of this unified dark fluid. From the definition of adiabatic sound speed
\begin{equation}
c_s^2=\left(\frac{\partial P}{\partial \rho}\right)_s=\frac{d P}{d\rho}=\rho\frac{dw}{d\rho}+w=\alpha,
\end{equation}
after integration,  one obtains its solution
\begin{equation}
w=\alpha-\frac{A}{\rho},
\end{equation}
where $A$ is the integration constant. The pressure $P=\alpha\rho-A$ can be obtained immediately. Assuming only gravitational interaction between energy components and considering the conservation of energy of dark fluid
\begin{equation}
\dot{\rho}_d+3H(1+w_d)\rho_d=0,
\end{equation}   
one has the energy density of dark fluid
\begin{equation}
\rho_d=\frac{A}{1+\alpha}+Ba^{-3(1+\alpha)},
\end{equation}
where $B$ is another integration constant. After the normalization of scale factor $a$ to $1$ today, $a_0=1$, one has  dark fluid energy density in the form
\begin{equation}
\rho_d=\rho_{d0}\left\{(1-B_s)+B_s a^{-3(1+\alpha)}\right\},\label{eq:rhod}
\end{equation}
where $B_s=B/\rho_{d0}$. In terms of $B_s$ and $\rho_{d0}$, one has $A=\rho_{d0}(1+\alpha)(1-B_s)$. For this unified dark fluid, one can do a decomposition, for example, $\rho_{dm}=\rho_{dm0}a^{-3}$ with usual power law and $\rho_{de}=\rho_{d}-\rho_{dm}$. If you like, you can do many kinds of decomposition. But, without any guide of physics principle, any kind of decomposition is non-proper. It has the possibility that it is just a whole component. And there does not exist any decomposition at all. So, in this paper, we will keep it as a whole unified dark fluid and do not implement decomposition any more. Then one can recast EoS of dark fluid $w_d$ into 
\begin{equation}
w_d=\alpha-\frac{(1+\alpha)(1-B_s)}{(1-B_s)+B_s a^{-3(1+\alpha)}}.
\end{equation}
in terms of $B_s$ and $\alpha$ which are model parameters in our story. It is interesting that the adiabatic sound speed $\alpha$ appears in the power law index of scale factor. It can be understood easily as follows. Since $c_s^2=\alpha$ has relation with perturbation propagates in fluid, it modifies the scaling relation of energy density with scale factor naturally. Staring at the expression of energy density of this dark fluid, one sees that a cosmological constant is recovered when $B_s=0$ and dark matter is obtained if $B_s=1$ and $\alpha=0$ are respected. To protect energy density from negativity, $0\le B_s\le 1$ is mandatory. In some sense, one can take this unified dark fluid energy density as a interpolation from 'dark matter' and cosmological constant. Just in this sense, it was dubbed as $\Lambda\alpha$CDM model by authors \cite{ref:lalphacdm}. Then one can find out that the EoS of this dark fluid is in the range of $w_d\in [-1,\alpha)$. We can show the evolutions of $w_d(a)$ with respect to the scale factor $a$ in Figure \ref{fig:wd}, where different values of model parameters $\alpha$ and $B_s$ are taken. From the left panel of Figure \ref{fig:wd}, one sees that this dark fluid is dark matter like in the early epoch when $\alpha$ approaches to zero in the limit $a\rightarrow 0$. \begin{widetext}
\begin{center}
\begin{figure}[htb]
\includegraphics[width=8.5cm]{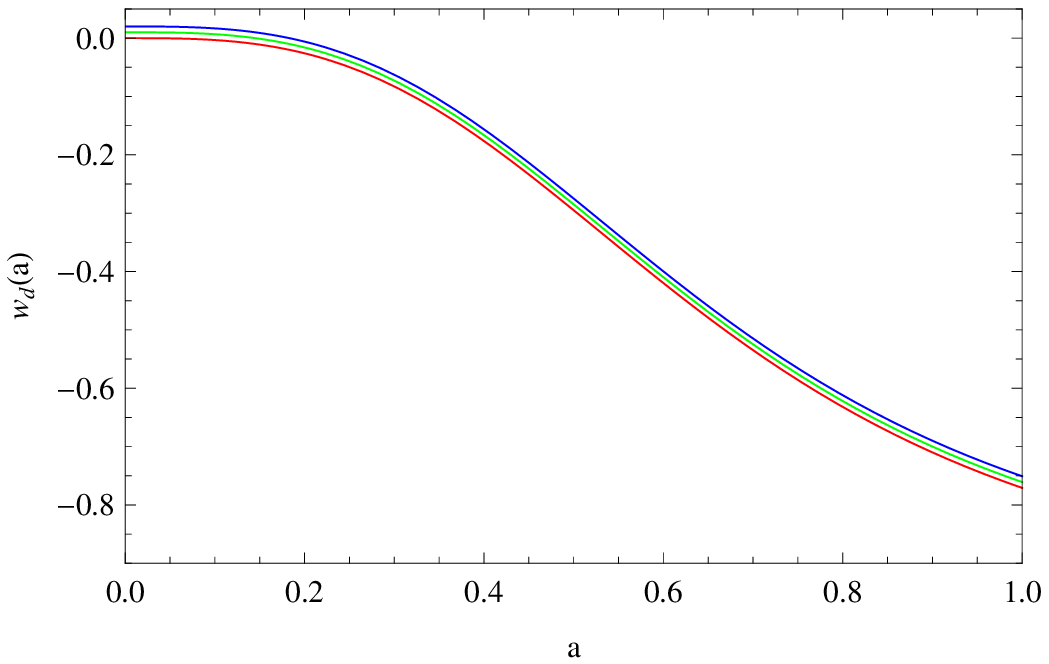}
\includegraphics[width=8.5cm]{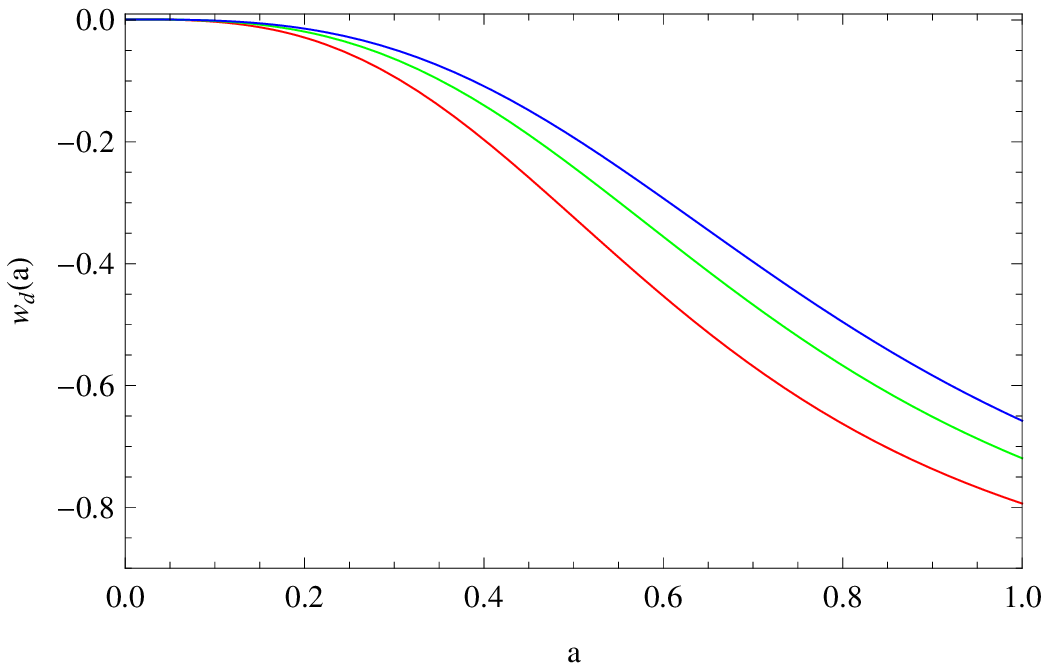}
\caption{The evolutions of $w_d(a)$ with respect to the scale factor $a$. In the left panel, $B_s=0.23$ is fixed. And the lines from the bottom to top correspond to $\alpha=0.01,0.02,0.04$. In the right panel, $\alpha=0.001$ is fixed. 
The lines from the bottom to top correspond to $B_s=0.2, 0.3, 0.4$.}\label{fig:wd}
\end{figure}
\end{center}
\end{widetext}

Consider a spatially flat FRW universe
\begin{equation}
ds^{2}=-dt^{2}+a^{2}(t)\left[dr^{2}+r^{2}(d\theta^{2}+\sin^{2}\theta d\phi^{2})\right],
\end{equation}
from the Einstein field equations, one has the Friedmann equation
\begin{widetext}
\begin{equation}
H^{2}=H^{2}_{0}\left\{(1-\Omega_{b}-\Omega_{r})\left[(1-B_{s})+B_{s}a^{-3(1+\alpha)}\right]+\Omega_{b}a^{-3}+\Omega_{r}a^{-4}\right\},
\end{equation}
\end{widetext}
where $H$ is the Hubble parameter with its current value $H_{0}=100h\text{km s}^{-1}\text{Mpc}^{-1}$, and $\Omega_{i}$ ($i=b,r$) are dimensionless energy parameters of baryon, radiation density respectively. 

Considering the perturbation in the synchronous gauge and the conservation of energy-momentum tensor $T^{\mu}_{\nu;\mu}=0$, one has the perturbation equations of density contrast and velocity divergence for dark fluid
\begin{eqnarray}
\dot{\delta}_d=-(1+w_d)(\theta_d+\frac{\dot{h}}{2})-3\mathcal{H}(c^{2}_{s}-w_d)\delta_d\\
\dot{\theta}_d=-\mathcal{H}(1-3c^{2}_{s})\theta_d+\frac{c^{2}_{s}}{1+w_d}k^{2}\delta_d-k^{2}\sigma_d
\end{eqnarray}
following the notations of Ma and Bertschinger \cite{ref:MB}. For the gauge ready formalism about the perturbation theory, please see \cite{ref:Hwang}. For the dark fluid in this paper, we assume the shear perturbation $\sigma_d=0$. 
In our calculations, the adiabatic initial conditions will be taken.

The linear and nonlinear instabilities were studied extensively in the context of unified dark fluid model \cite{ref:stability}. For the linear perturbations, the adiabatic sound speed of the unified dark fluid $c^2_s=\alpha$ is kept in the range of $0$ to $1$ in our model, so the linear instability is avoided. The averaging problem is involved when perturbations become nonlinear. The problem coms from the fact that $\langle p\rangle\neq p(\langle\rho\rangle)$ for unified dark fluid models. For example, in the generalized Chaplygin gas (GCG) model \cite{ref:GCG}
\begin{equation}
p=-\frac{A}{\rho^\beta}
\end{equation}
where $A$ and $\beta$ are positive model parameters in GCG model, one can check that 
\begin{equation}
\langle p\rangle=-\langle A/\rho^\beta\rangle\neq -A/\langle\rho\rangle^\beta=p(\langle\rho\rangle),
\end{equation}
in the case of $\beta\neq 0$. However, it is not the issue in our model for the linear relations between $p$ and $\rho$, i.e. $p=\alpha\rho-A$. It is a feature of our unified dark fluid model.

\subsection{Scalar Fields}

In \cite{ref:stability}, the authors discussed the relations of GCG with scalar fields. Then what about the unified dark fluid with a constant adiabatic sound speed? In this subsection, we will derive its Lagrangian density following the procedure of \cite{ref:stability}. For a single scalar field, the action is generalized to the form \cite{ref:stability,ref:k-essence}
\begin{equation}
S_\phi=\int d^4 x\sqrt{-g}\mathcal{L}(X,\phi),
\end{equation}
where $\mathcal{L}$ is the Lagrangian density, $\phi$ is a real scalar field, and $X$ is its kinetic term $X=-\nabla^\mu\nabla_\mu\phi/2$. Then the energy density and pressure are given as
\begin{equation}
\rho=2Xp_{,X}-p,\quad p=\mathcal{L}(X,\phi).\label{eq:scalar}
\end{equation}
To get the Lagrangian density is to solve the first differential equation of Eq. (\ref{eq:scalar}) in terms of $X$. For our unified dark fluid model, one has
\begin{equation}
2\alpha X p_{,X}-(1+\alpha)p-A=0.
\end{equation}
For $\alpha=0$, the solution is $p(X)=-A$. For $\alpha\neq 0$, 
one has the solution
\begin{equation}
p=\alpha X^{(1+\alpha)/(2\alpha)}-A/(1+\alpha).
\end{equation}
In terms of $X$, the energy density is given as
\begin{equation}
\rho=A/(1+\alpha)+X^{(1+\alpha)/(2\alpha)}.
\end{equation}
Thus the Lagrangian density is 
\begin{equation}
\mathcal{L}(X)=p(X)=\alpha X^{(1+\alpha)/(2\alpha)}-A/(1+\alpha)
\end{equation}
with $0<X<1/2$. One can check that the adiabatic sound speed is
\begin{equation}
c^2_s\equiv \frac{p_{,X}}{\rho_{,X}}=\alpha.
\end{equation}
Here please note that it is the adiabatic sound speed not the sound speed $c^2_{s\phi}=1$ which is defined in the scalar filed rest frame, i.e $\delta\phi|_{rf}=0$ \cite{ref:soundspeed}.

\section{Constraint method and results}\label{sec:method}

\subsection{Implications on CMB anisotropy}

Before going to the issue of cosmic observational constraint, we would like to show some implications on CMB anisotropy in this unified dark fluid model. In Figure \ref{fig:cmb}, we illustrate how the CMB temperature anisotropies are characterized by different values of $\alpha$, by choosing different values of $B_s$ with the other cosmological parameters fixed.
\begin{widetext}
\begin{center}
\begin{figure}[htb]
\includegraphics[width=8.5cm]{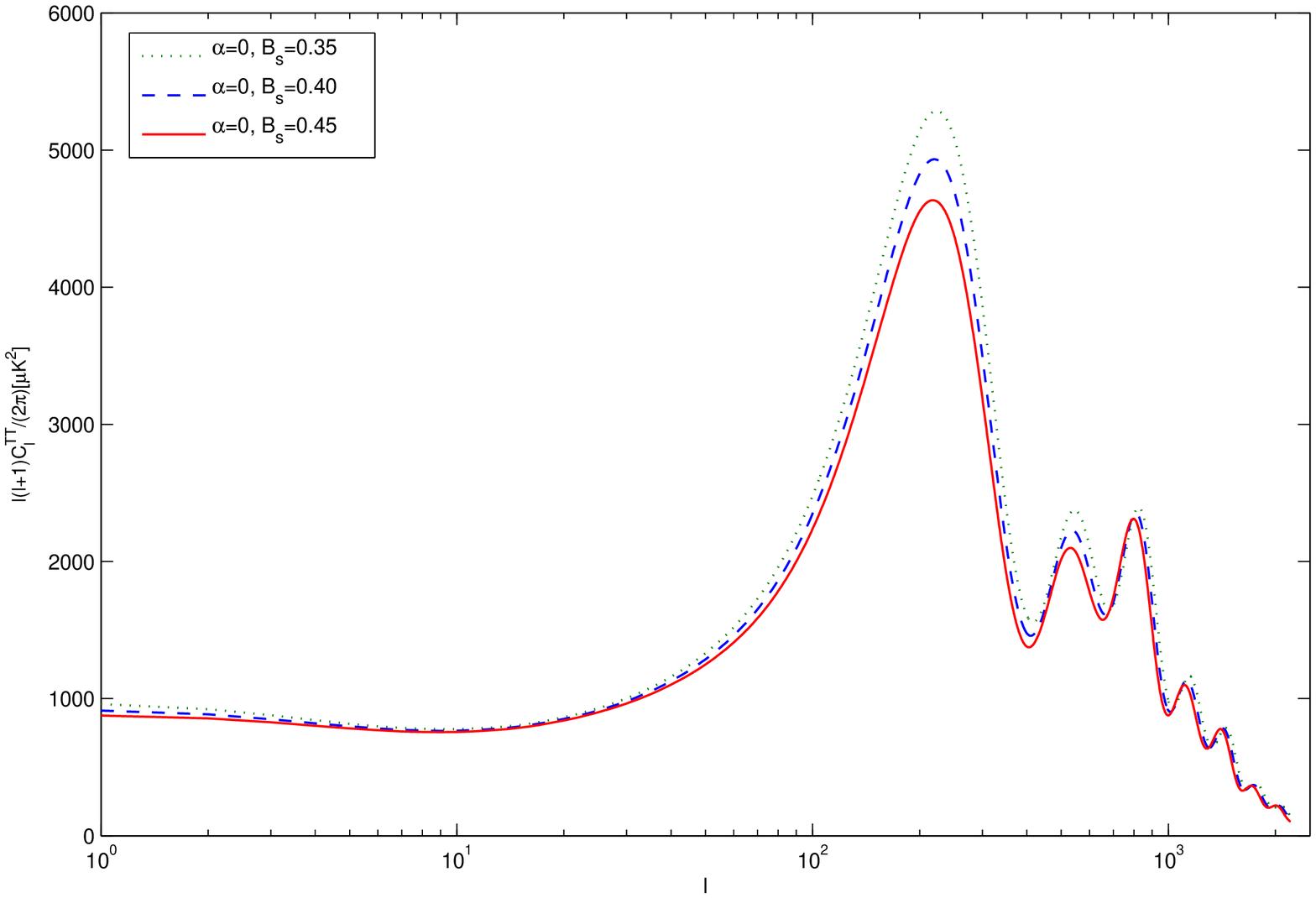}
\includegraphics[width=8.5cm]{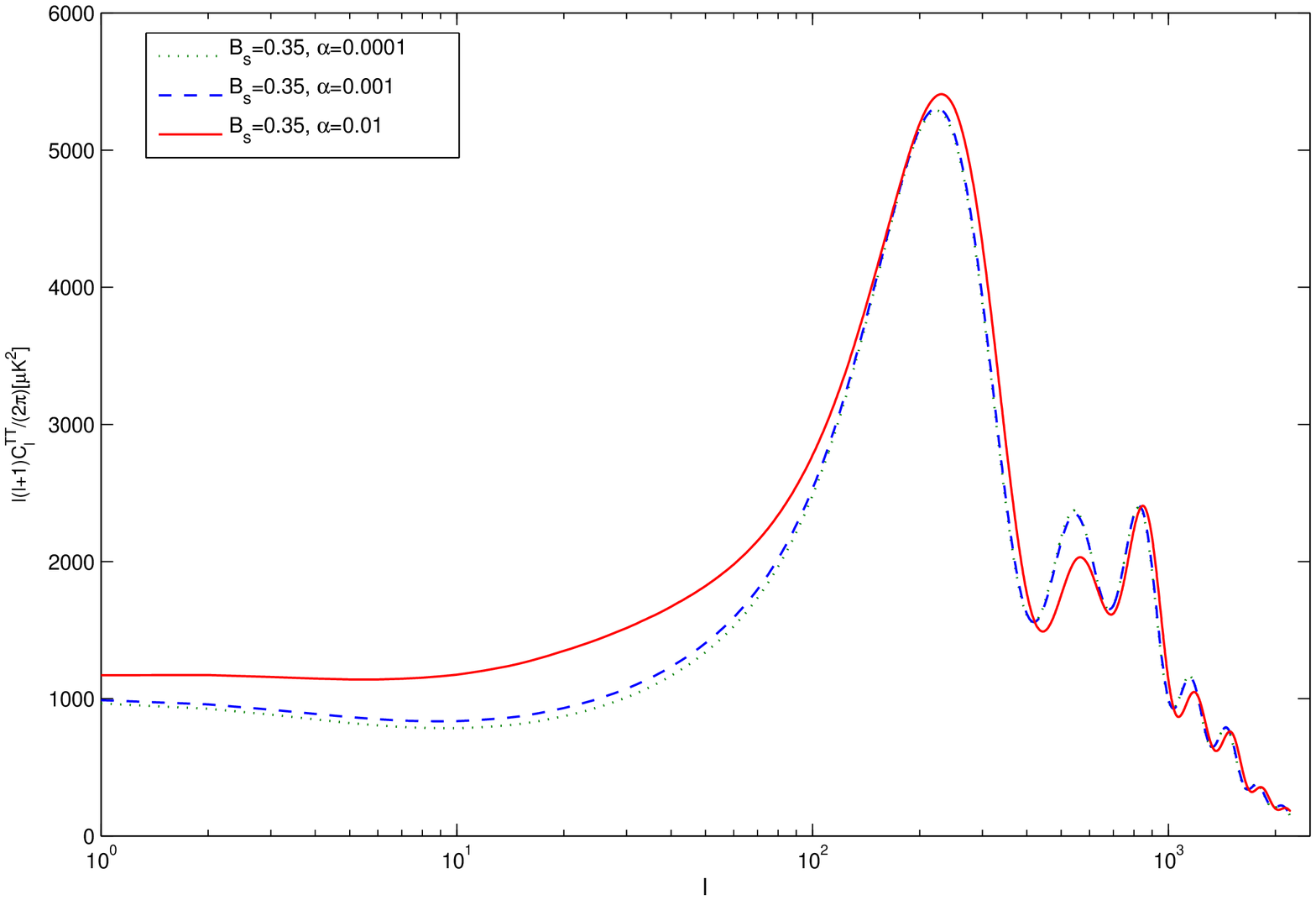}
\caption{The effects on CMB temperature power spectra of a unified dark fluid with constant adiabatic sound speed $c_s^2=\alpha$ characterized by different values of $\alpha$ and $B_s$. On the left panel, the fixed value of $\alpha=0$ is adopted. The bottom red solid line is with $B_s=0.45$ and top green dotted line is with $B_s=0.35$, and the center one is with $B_s=0.40$. On the right panel, $B_s=0.35$ is fixed. The lines take the values of $\alpha=0.0001, 0.001,0.01$  from bottom to the top.}\label{fig:cmb}
\end{figure}
\end{center}
\end{widetext}
From the left panel of Figure \ref{fig:cmb}, as is expected $B_s$ corresponds to an effective dimensionless energy density of dark matter at early epoch. So, increasing the values of $B_s$ will make the equality of matter and radiation earlier, then the sound horizon is decreased. As a result, the first peak is depressed. The variation of the second peak is from the variation of ratio of baryon and effective dark matter. The right panel of Figure \ref{fig:cmb} shows the effects of $\alpha$ on the CMB power spectra. As shown in Eq. (\ref{eq:rhod}), the values of $\alpha$ describe the possible deviation from standard evolution scaling law $a^{-3}$ of effective dark matter. It affects the CMB power spectra at large scales $l < 100$ via Integrated Sachs-Wolfe (ISW) effect due to the evolution of gravitational potential. At small scales, the effect comes from the different scaling of effective dark matter evolution. From the Figure \ref{fig:cmb}, one can read that the CMB power spectra are sensitive to the values of $\alpha$ and $B_s$. Then the full CMB information will be helpful to constrain the model parameter space.

\subsection{Method and data points}

To obtain the parameter space, the observational constraints are performed by using Markov Chain Monte Carlo (MCMC) method. We modified publicly available cosmoMC package \cite{ref:MCMC} to include the dark fluid perturbation in the CAMB \cite{ref:CAMB} code which is used to calculate the theoretical CMB power spectrum. The following $7$-dimensional parameter space  is adopted
\begin{equation}
P\equiv\{\omega_{b},\Theta_{S},\tau, \alpha,B_{s},n_{s},\log[10^{10}A_{s}]\}
\end{equation}
where $\omega_{b}=\Omega_{b}h^{2}$ is the physical baryon density, $\Theta_{S}$ (multiplied by $100$) is the ration of the sound horizon and angular diameter distance, $\tau$ is the optical depth, $\alpha$ and $B_{s}$ are two newly added model parameters related to dark fluid, $n_{s}$ is scalar spectral index, $A_{s}$ is the amplitude of of the initial power spectrum. The pivot scale of the initial scalar power spectrum $k_{s0}=0.05\text{Mpc}^{-1}$ is used in this paper. The following priors to model parameters are adopted: $\omega_{b}\in[0.005,0.1]$, $\Theta_{S}\in[0.5,10]$, $\tau\in[0.01,0.8]$, $\alpha\in[0,1]$, $B_{s}\in[0,1]$, $n_{s}\in[0.5,1.5]$, $\log[10^{10}A_{s}]\in[2.7, 4]$. Furthermore, the hard coded prior on the comic age $10\text{Gyr}<t_{0}<\text{20Gyr}$ is also imposed. Also, the physical baryon density $\omega_{b}=0.022\pm0.002$ \cite{ref:bbn} from big bang nucleosynthesis and new Hubble constant $H_{0}=74.2\pm3.6\text{kms}^{-1}\text{Mpc}^{-1}$ \cite{ref:hubble} are adopted.

To get the distribution of parameters, we calculate the total likelihood $\mathcal{L} \propto e^{-\chi^{2}/2}$, where $\chi^{2}$ is given as
\begin{equation}
\chi^{2}=\chi^{2}_{CMB}+\chi^{2}_{BAO}+\chi^{2}_{SN}.
\end{equation}
The $557$ Union2 data \cite{ref:Union2} with systematic errors and BAO \cite{ref:BAO} are used to constrain the background evolution, for the detailed description please see Refs. \cite{ref:Xu}. SN Ia is used as standard candle. And BAO is used as standard ruler. At the last scattering of CMB radiation, the acoustic oscillation in the baryon-photon fluid was frozen and imprinted their signature on the matter distribution. The characterized scale of BAO in the observed galaxy power spectrum is determined by the comoving sound horizon at drag epoch $z_d$ which is shortly after photon decoupling. To calculate $r_{s}(z_{d})$, one needs to know the redshift $z_d$ at decoupling epoch and its corresponding sound horizon. In our story, the usual fitting formula \cite{ref:EH} can not be used for its viability under the conditions $\rho_{b}\propto a^{-3}$ and $\rho_{c}\propto a^{-3}$. So, to use the BAO information, we obtain the baryon drag epoch redshift $z_d$ numerically from the following integration \cite{ref:Hamann}
\begin{eqnarray}
\tau(\eta_d)&\equiv& \int_{\eta}^{\eta_0}d\eta'\dot{\tau}_d\nonumber\\
&=&\int_0^{z_d}dz\frac{d\eta}{da}\frac{x_e(z)\sigma_T}{R}=1
\end{eqnarray}   
where $R=3\rho_{b}/4\rho_{\gamma}$, $\sigma_T$ is the Thomson cross-section and $x_e(z)$ is the fraction of free electrons. Then the sound horizon is
\begin{equation}
r_{s}(z_{d})=\int_{0}^{\eta(z_{d})}d\eta c_{s}(1+z).
\end{equation}   
where $c_s=1/\sqrt{3(1+R)}$ is the sound speed. Also, to obtain unbiased parameter and error estimates, we use the substitution \cite{ref:Hamann}
\begin{equation}
d_z\rightarrow d_z\frac{\hat{r}_s(\tilde{z}_d)}{\hat{r}_s(z_d)}r_s(z_d),
\end{equation}
where $d_z=r_s(\tilde{z}_d)/D_V(z)$, $\hat{r}_s$ is evaluated for the fiducial cosmology of Ref. \cite{ref:BAO}, and $\tilde{z}_d$ is redshift of drag epoch obtained by using the fitting formula \cite{ref:EH} for the fiducial cosmology. Here $D_V(z)=[(1+z)^2D^2_Acz/H(z)]^{1/3}$ is the 'volume distance' with the angular diameter distance $D_A$. In this paper, for BAO information, the SDSS data points from \cite{ref:BAO} are used. For CMB data set, the temperature power spectrum from WMAP $7$-year data \cite{ref:wmap7} are employed as dynamic constraint.

\subsection{Fitting Results and discussion}

After running $8$ independent chains and checking the convergence to stop sampling when $R-1$ is of order $0.01$, the global fitting results are summarized in Table \ref{tab:results} and Figure \ref{fig:contour}. We find that the minimum $\chi^2$ is $\chi^2_{min}=8009.424$ which is a little bigger than that for $\Lambda$CDM model $8009.116$ for the same cosmic observational data sets combination. From the Table \ref{tab:results}, one can see that small values of $c_s^2=\alpha$ are favored in this unified dark fluid model when the WMAP $7$-year observation, SN Union2 and BAO data points are used as cosmic constraint.
It is worth to mention that our result is consistent with that reported in Ref. \cite{ref:lalphacdm} where a relative loose constraint $\alpha=0.01\pm0.02$ was given. Correspondingly, we plot the CMB power spectrum for the mean values estimated from MCMC analysis in Figure \ref{fig:mean}, where the observational data points of $7$-year WMAP with uncertainties are also included. As comparisons, the CMB temperature power spectrum for $\Lambda$CDM with same cosmic observational data sets combination and results from \cite{ref:wmap7} are also shown. One can see that the CMB power spectrum for this unified dark fluid model is well inside the error bars of the binned measurements from WMAP $7$-year results and almost matches to $\Lambda$CDM model. This strongly implies that current cosmic observational data combinations of SN Union2, BAO and WMAP $7$-year can not almost discriminate a unified dark fluid with constant adiabatic sound speed from $\Lambda$CDM model, though $\Lambda$CDM model is still slightly favored.  
\begin{table}[tbh]
\begin{center}
\begin{tabular}{cc}
\hline\hline Prameters&Mean with errors\\ \hline
$\Omega_b h^2$ & $    0.0225_{-    0.000499-    0.000992}^{+    0.000504+    0.000979}$ \\
$\theta$ & $  1.0484_{-    0.00245-    0.00487}^{+    0.00246+    0.00491}$ \\
$\tau$ & $    0.0874_{-    0.00701    0.0225}^{+    0.00607+    0.0243}$ \\
$\alpha$ & $    0.000487_{-    0.000487-    0.000487}^{+    0.000117+    0.000728}$ \\
$B_s$ & $    0.229_{-    0.0133-    0.0249}^{+    0.0134+    0.0274}$ \\
$n_s$ & $    0.975_{-    0.0131-    0.0249}^{+    0.0132+    0.0265}$ \\
$\log[10^{10} A_s]$ & $    3.081_{-    0.0327-    0.0616}^{+    0.0331+    0.0663}$ \\
$\Omega_{d}$ & $    0.956_{-    0.00150-    0.00300}^{+    0.00152+    0.00298}$ \\
$Age/Gyr$ & $   13.699_{-    0.1085-    0.215}^{+    0.108+    0.210}$ \\
$\Omega_b$ & $    0.044_{-    0.00152-    0.00298}^{+    0.00150+    0.00300}$ \\
$z_{re}$ & $   10.523_{-    1.1618-    2.299}^{+    1.1948+    2.393}$ \\
$H_0$ & $   71.341_{-    1.268-    2.478}^{+    1.278+    2.573}$ \\
\hline\hline
\end{tabular}
\caption{The mean values of model parameters with $1\sigma$ and $2\sigma$ errors, where WMAP $7$-year, SN Union2 and BAO data sets are used.}\label{tab:results}
\end{center}
\end{table}
\begin{widetext}
\begin{center}
\begin{figure}[htb]
\includegraphics[width=17cm]{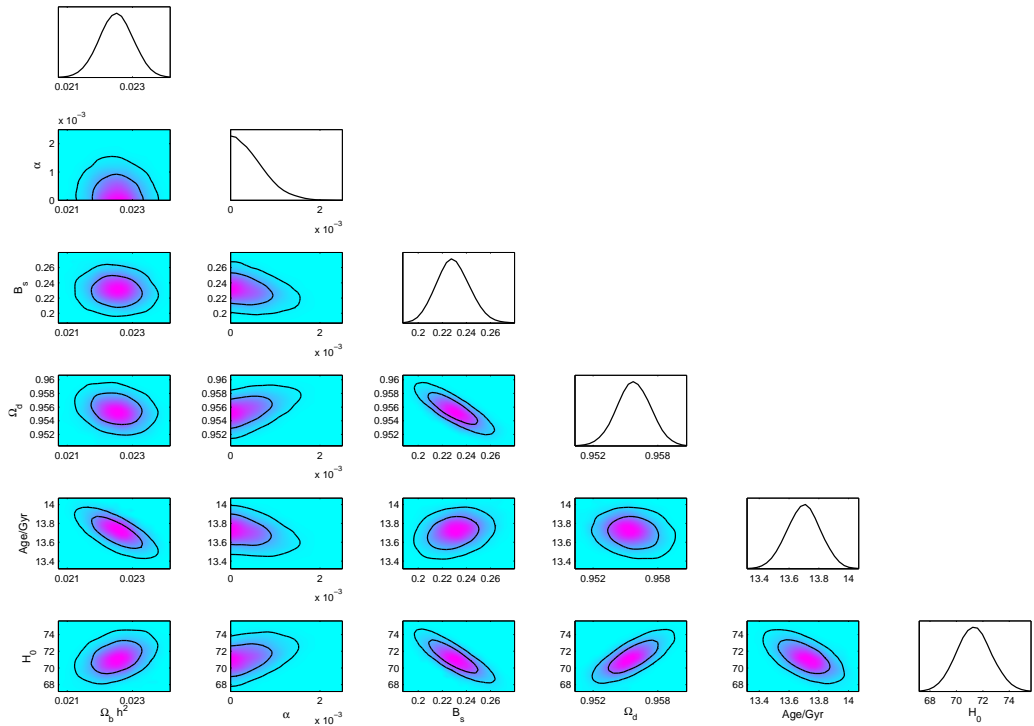}
\caption{The 1D marginalized distribution on individual parameters and 2D contours  with $68\%$ C.L. and $95\%$ C.L. by using CMB+BAO+SN data points. The shade regions show the mean likelihood of the samples.}\label{fig:contour}
\end{figure}
\end{center}
\end{widetext}
\begin{widetext}
\begin{center}
\begin{figure}[htb]
\includegraphics[width=17cm]{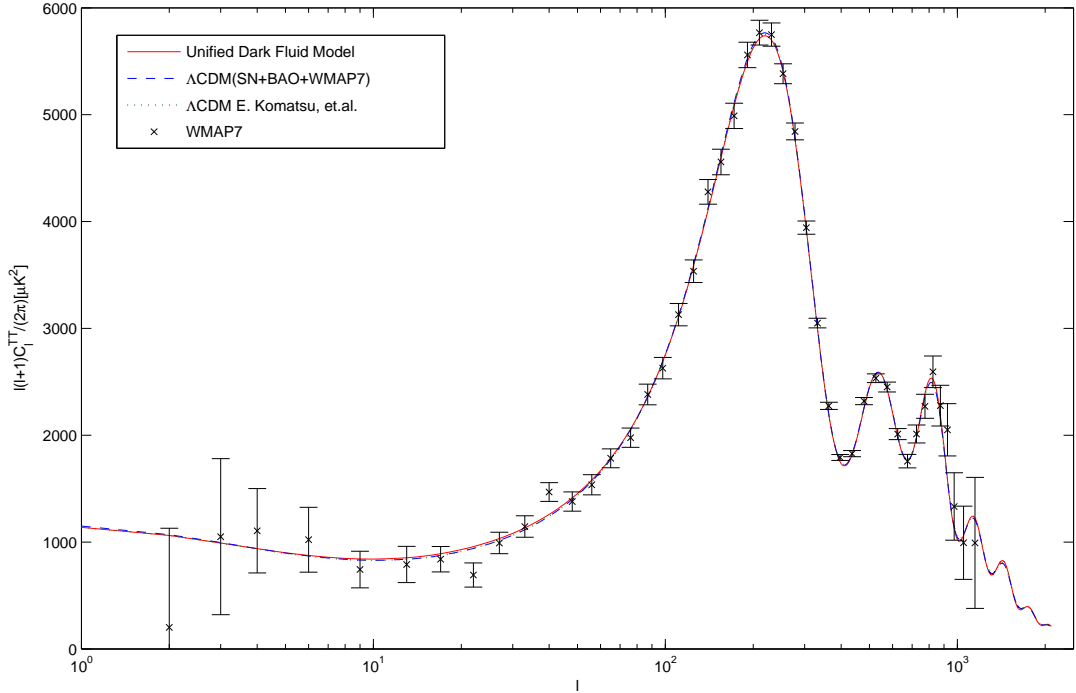}
\caption{The CMB $C^{TT}_l$ power spectrum v.s. multiple moment $l$, where the black dots with error
bars denote the observed data with their corresponding uncertainties from WMAP $7$-year results, the red solid line is for the unified dark fluid model with mean values as shown in Table \ref{tab:results}, the blue dashed line is for $\Lambda$CDM model with mean values for the same data points combination. And the green doted line is for $\Lambda$CDM model with mean values taken from \cite{ref:wmap7} with WMAP+BAO+$H_0$ constraint results.}\label{fig:mean}
\end{figure}
\end{center}
\end{widetext}

Up to now, by using the geometric informations from SN, BAO and CMB data sets, one has obtained a relative tight constraint to the model parameter space. However, for any real Universe model, one needs to consider the large scale structure formation. That is to say, the dynamic evolution information from galaxy formation would be employed. It is not an easy issue for unified dark fluid model. Here, we just show that the smaller values of $\alpha < 10^{-5}$ are required for matching the matter (baryon) power spectra data points from SDSS DR7 \cite{ref:sdssdr7}, please see Figure \ref{fig:pk}. It means that when the dynamic information is used, for example large scale structure formation, a tighter constraint will be obtained. For the constraint to this unified dark fluid model from large structure information, we leave it for the future work.  
\begin{widetext}
\begin{center}
\begin{figure}[htb]
\includegraphics[width=16cm]{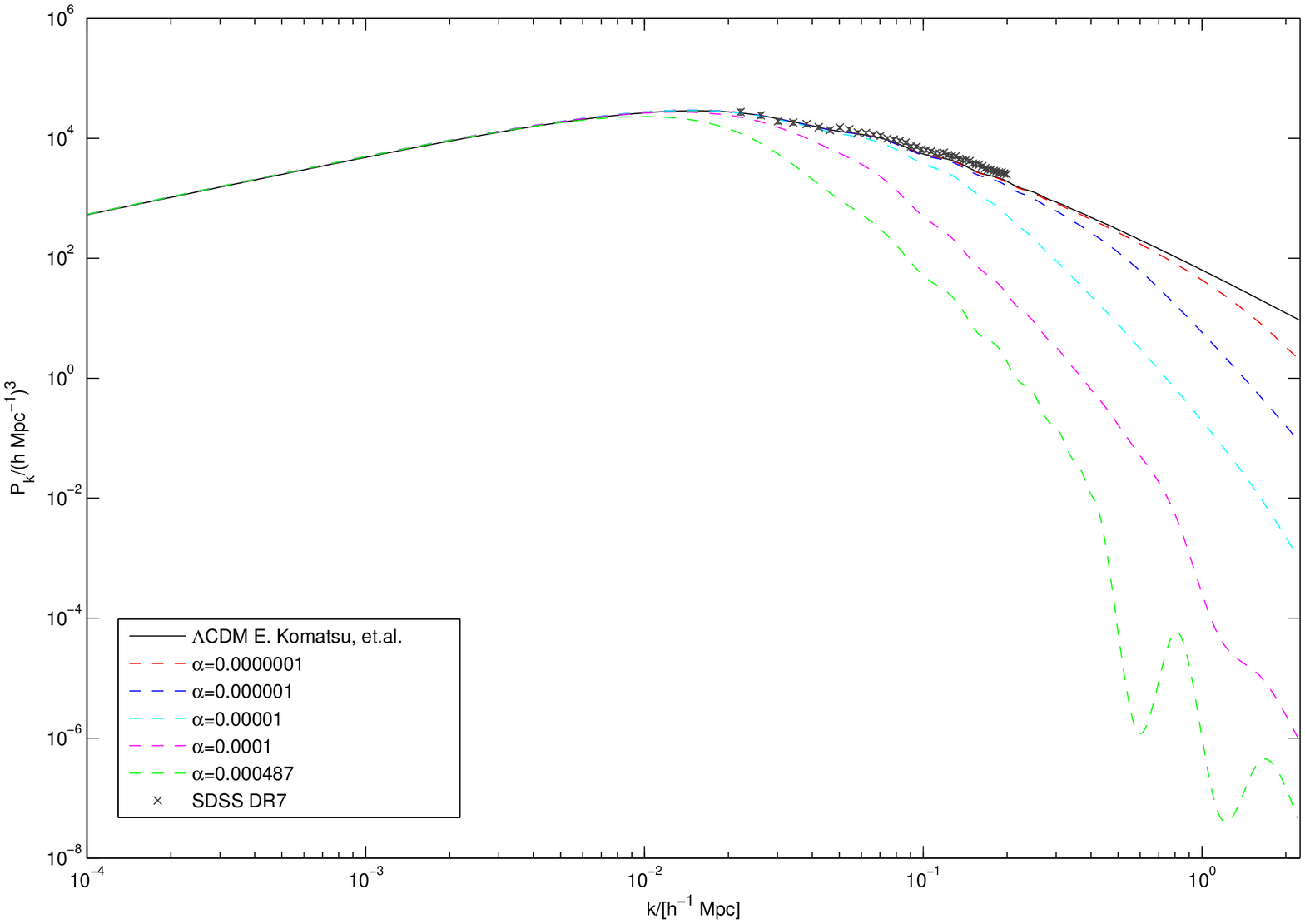}
\caption{The matter (baryon) power spectra v.s. wave number $k/h$, where the grey dots with error
bars denote the observed data with their corresponding uncertainties from SDSS DR7, the solid black line is for  $\Lambda$CDM model with mean values taken from \cite{ref:wmap7} for WMAP+BAO+$H_0$ constraint results, the dashed lines for unified dark fluid model from the top to the bottom correspond to $\alpha=0.0000001, 0.000001, 0.00001, 0.0001, 0.000487$ respectively with mean values as shown in Table \ref{tab:results}. }\label{fig:pk}
\end{figure}
\end{center}
\end{widetext}

\section{Summary} \label{ref:conclusion} 

In this paper, we have studied a unified dark fluid model with constant adiabatic sound speed $c^2_s=\alpha$. This property is called dark degeneracy. By using MCMC method with current available cosmic observational data sets combinations of SN Unioin2, BAO and WMAP $7$-year results, we find that small values of $\alpha$ are favored in this unified dark fluid model as shown in Table \ref{tab:results} and Figure \ref{fig:contour}. For small values of $\alpha$, this unified dark fluid model approaches to $\Lambda$CDM model closely. Our analysis shows that $\Lambda$CDM model is still slightly favored when SN Union2, BAO and WMAP $7$-year observational results as useful constraint. But, there exists the possibility that the unknown above $90\%$ content of our universe is made of a unified dark fluid, the so-called dark degeneracy property. To break the degeneracy, more cosmic observational data sets would be included. We also show that smaller values of $\alpha <10^{-5}$ are required to match the matter (baryon) power spectra from SDSS DR7. We expect our study can shed light on the understanding the dark side of our Universe.

\acknowledgements{We thank an anonymous referee for helpful improvement of this paper. Xu thanks Dr. Zhiqi Huang for useful discussion about matter power spectra. L. Xu's work is supported by the Fundamental Research Funds for the Central Universities (DUT10LK31) and (DUT11LK39). H. Noh's work is supported by Mid-career Research Program through National Research Foundation funded by the MEST (No. 2010-0000302).}

\end{document}